\def\be{\begin{equation}}
\def\ee{\end{equation}}
\def\bea{\begin{eqnarray}}
\def\eea{\end{eqnarray}}
\documentclass[aps,prl,twocolumn,superscriptaddress,amsmath,amssymb,showpacs,showkeys]{revtex4}
\usepackage{bm}
\usepackage{graphicx}
\usepackage{epsfig}
\usepackage{psfrag}
\usepackage{psfrag}
\usepackage{epstopdf}
\usepackage{subfig}
\begin{document}

\title{Heisenberg Symmetry and Collective Modes of One Dimensional Unitary Correlated Fermions}

\author{Kumar Abhinav}
\email{kumarabhinav@iiserkol.ac.in}
\affiliation {Indian Institute of Science Education and Research  Kolkata, Mohanpur-741246, India}
\author{Chandrasekhar Bhamidipati}
\email{chandrasekhar@iitbbs.ac.in}
\affiliation{Indian Institute of Technology Bhubaneswar, Bhubaneswar-751013, India}
\author{Vivek M. Vyas}
\email{vivekmv@imsc.res.in}
\affiliation{Institute of Mathematical Sciences, Tharamani, Chennai-600113, India}
\author{Prasanta K. Panigrahi}
\email{pprasanta@iiserkol.ac.in}
\affiliation {Indian Institute of Science Education and Research  Kolkata, Mohanpur-741246, India}

\date{\today}

\begin{abstract}
The correlated fermionic many-particle system, near infinite scattering length, reveals
an underlying Heisenberg symmetry in \textit{one} dimension, as compared to an $SO(2,1)$
symmetry in two dimensions. This facilitates an exact map from the interacting to the
non-interacting system, both with and without a harmonic trap, and explains the short-distance
scaling behavior of the wave-function. Taking advantage of the phenomenological Calogero-Sutherland-type
interaction, motivated by the density functional approach, we connect the ground-state energy
shift, to many-body correlation effect. For the excited states, modes at integral
values of the harmonic frequency $\omega$, are predicted in one dimension, in contrast to
the breathing modes with frequency $2\omega$ in two dimensions.
\end{abstract}
\maketitle

The behavior of interacting systems in one spatial dimension (1-D) \cite{F1}, with increased
correlation and fluctuations \cite{F3} shows significant differences with their
higher dimensional counterparts, which is particularly predominant in the {\it unitary} regime \cite{F2}, characterized
by infinite scattering length $a$.  The scattering cross-section
in 1-D behaves as \cite{Griffiths},

\be
\sigma(k)\propto \frac{1}{1+k^2a^2},\nonumber
\ee
vanishing at unitarity. In contrast, for 3-D with $\sigma(k)\propto a^2/\left[1+k^2a^2\right]\xrightarrow{a\rightarrow\infty}1/k^2$, the behavior is solely determined by the relative momentum, $k=\vert\vec{k}\vert$ \cite{Pitaevskiireview}. 
The latter case has been realized recently
in dilute Fermi gases \cite{Eagles}, wherein unitarity was achieved through Feshbach resonance \cite{Pitaevskiireview,N1}, by tuning
the relative scattering energy close to that of a bound state by an external magnetic field, leading to a
substantial resonant scattering \cite{N2}. A hallmark feature of such systems
is that the pairing correlations have a range shorter than the de Broglie wavelength, leading to significantly
novel behavior \cite{Pethick,Randeria}. 
\paragraph*{}The lack of a small parameter at unitarity
poses a challenging problem in understanding the structure of many-body wave-functions
and the associated bound states. In 3-D, the only available length scale in the
ground state is the inter particle spacing, $n^{-1/3}$, where $n$ is the density, 
since $a$ must drop out from all physical observables. This ensures {\it Universality} 
for both bosons and fermions, with only relevant energy scale being the Fermi
energy $\epsilon_F$. The ground-state energy $\epsilon_0$ of the unitary fermions has to then scale as
$\xi\epsilon_F$ \cite{three}, where the parameter $\xi$ does not depend on the specific form of the
short range potential. In the molecule formation regime, with singular wave-functions,
it is same for all potentials, having an $s$-wave bound state. Understanding the origin of such scaling behavior is of prime 
importance, as it appears in various branches of physics, such as the gravity side of AdS/CFT
correspondence \cite{N002}.
\paragraph*{}From the theoretical view point, a careful inspection~\cite{Nussinov}
of the short distance behavior of the two-body wave-function  $\psi$, in the
unitary regime, shows a scaling behavior~\cite{Werner},

\be \label{scalingregime}
Q\psi=\gamma\psi,~~~Q=\sum_{i=1}^N\mathbf{r}_i\cdot \mathbf{\nabla}_i ,
\ee
where $\gamma$ is the scaling exponent, brought out by the Euler operator $Q$.
Thus, at unitarity, the short-range behavior of the wave-function
is mononomial, with each coordinate having same exponent $\gamma$,
and none other, which represents Universal inter-particle correlation. 
A scaling Hamiltonian of the form $H\rightarrow\lambda^{-2}H$,
corresponding to the scale transformation $x\rightarrow \lambda {x}$, possesses  
wave-functions transforming as $\Psi(\lambda x) \rightarrow \lambda^\gamma \Psi(x)$ \cite{Werner}.
The potential has to scale as $V(\lambda x) = \lambda^{-2}V(x)$, which, for unitarity,
additionally needs to be short-ranged corresponding to infinite scattering length \cite{Son}.
\paragraph*{}The scaling Hamiltonian has been shown to have $SO(2,1)$ Lorentz symmetry
\cite{Pitaevskii} in a harmonic trap, leading to Universal scaling of 
the ground-state energy \cite{Stan}:

\be
\epsilon_0 = \left(\gamma + \frac{N}{2}d\right)\hbar\omega, \label{F1}
\ee
for {\it any} short-range interaction in $d$-dimensions. Such weakly interacting particles can
show several new features. In 2-D, such a system can host additional breathing modes with
universal frequency $2\omega$ \cite{Pitaevskii,Pitaevskii1,Benjamin}. The physical origin
of the `shift' part $\gamma\hbar\omega$ is of prime importance. Interestingly, this shift
corresponds to the short-ranged interaction, that identically yields scaling behavior with
exponent $\gamma$.  
%%%%%%%%%%%%%%%%%%%%%%%%%%%%%%%%%%%%%%%%%%%%%%%%%%%%%%%%%%%%%%%%%%%%%%%%%%%%%%%%
\paragraph*{}The goal of this paper is to investigate the unitary Fermi gas in 1-D. We confine
ourselves to the correlated systems away from the domain of molecule formation. It is observed that 
the underlying algebraic structure is of Heisenberg-type, as compared to the $SO(2,1)$ symmetry
of 2-D. This allows for an exact map from the interacting to the non-interacting system, both in
presence and absence of a harmonic trap and explicates the short-distance scaling behavior of
the wave-function. The Heisenberg symmetry depicts the harmonic motion of the center of mass
with frequency $\omega$ \cite{Kohn}, the radial $2\omega$ mode obeying $SO(2,1)$ algebra, together  
with angular modes having frequency in integral multiples of same $\omega$ There have been several works
in recent literature, which suggest a number of universal features connecting interacting and
non-interacting regimes \cite{pkp1,Werner,Blume,Son:2008ye,N01,N02}. Motivated by the
results of density functional treatment of unitary fermions \cite{DFCS}, the phenomenological
Calogero-Sutherland (C-S) type inverse-square interaction \cite{calogero} is considered, 
to illustrate the effect of correlation and scaling to the ground-state
energy-shift $\gamma\hbar\omega$. The scaling exponent $\gamma$ will be
related to pair-wise correlation, arising from Jastrow-type wave-functions, away from the
molecule-formation regime. 
%%%%%%%%%%%%%%%%%%%%%%%%%%%%%%%%%%%%%%%%%%%%%%%%%%%%%%%%%%%%%%%%%%%%%%%%%%%%%%%%%%%%%%%%%%
\paragraph*{}As noted earlier, one-dimensional systems are special as in 1-D, fermionic
fields can be written in terms of bosonic variables, and interacting bosonic systems like
Tonk-Girardeau gas show fermionic behavior \cite{N3}. The vanishing of $\sigma(k)$ in
1-D unitarity systems, with short-range interaction, implies {\it free} particles with
arbitrary energy. Furthermore, there is no angular degrees of freedom. It is, therefore,
expected that collective modes in 1-D will show unique characteristics. Particles are
expected to resonate {\it independently} under external periodic influence. %Further, the
%reduction of center-of-mass degree of freedom in 1-D, rules-out the applicability of
%Kohn's theorem \cite{Kohn}.  
%%%%%%%%%%%%%%%%%%%%%%%%%%%%%%%%%%%%%%%%%%%%%%%%%%%%%%%%%%%%%%%%%%%%%%%%%%%%%%%%%%%%%%%%%%%%%%% 
\paragraph*{}In the presence of a generic potential $V(\{x_i\})$, scaling like the kinetic energy,
a formal equivalence, between the interacting system and the non-interacting one, can be established
exactly. Consider the $N$-body Hamiltonian in a harmonic trap,

\be
H_T=-\frac{1}{2}\sum_i\frac{d^2}{dx_i^2}+V(\{x_i\})+\frac{1}{2}\omega^2\sum_ix_i^2,~~~\hbar=1=m.\nonumber
\ee 
A similarity transformation with $\exp\left\{(\omega/2)\sum_ix_i^2\right\}$, decouples the
center of mass motion \cite{Pitaevskii,N5}, %implying the obtained mode to be of breathing nature,
leaving,

\bea
\bar{H}&=&\exp\left(\frac{\omega}{2}\sum_ix_i^2\right)H_T\exp\left(-\frac{\omega}{2}\sum_ix_i^2\right)\nonumber\\
&=&{\cal A}+\omega\sum_iD_i+\frac{\omega}{2}N;\label{Hbar}\\
{\cal A}&:=&-\frac{1}{2}\sum_i\frac{d^2}{dx_i^2}+V(\{x_i\}).\nonumber
\eea
The identity \cite{pkp2},

\be
\sum_ix_i^2=NR^2+r^2,\label{Identity}
\ee
separates center of mass and `radial' coordinates, $R=(1/N)\sum_ix_i$ and $r^2=(1/N)\sum_{i<j}\left(x_i-x_j\right)^2$
respectively. Thus, $\bar{H}$ represents a disjoint Hilbert space sector of the initial system
($H_T$), representing the radial and $N-2$ `angular' degrees of freedom in the configuration space. These
excitations together will be shown to arise from the Heisenberg algebra. Physically, the
long-range aspects, imposed by the harmonic trap, are removed by this `Gaussian' 
transformation, leaving-out the short-range ones, represented through $V(\{x_i\})$. Another such
transformation with $\exp\{-{\cal A}/2\omega\}$ removes the effect of local correlation, leaving out
only the `Universal' scaling nature, through the {\it non-interacting} Hamiltonian,

\bea
H_S&=&\exp\left(-\frac{1}{2\omega}{\cal A}\right)\bar{H}\exp\left(\frac{1}{2\omega}{\cal A}\right)\nonumber\\
&\equiv&\omega\sum_iD_i+\frac{\omega}{2}N.\label{Euler}%{\cal H}+\omega\sum_iD_i+\frac{\omega}{2}N,
\eea
Here $D_i=x_i\frac{d}{dx_i}$ is the Euler operator, bringing out the scaling nature of the system,
owing to $V(\{x_i\})$. At the level of $H_S$, the eigenfunctions are symmetric, characterized
by their degree. Particularly, identifying the canonical pair of operators
$\left(x_i,\frac{d}{dx_i}\right)$, the ground-state can be defined as $\frac{d}{dx_i}\phi_0 = 0$. 
The 1-D space allows
for symmetric excited states of the form $\prod_l^N (x_i)^{n_l}$, formed by Cartesian particle
coordinates $x_i$. The
resultant spectrum $\omega\sum_ln_l+E_0$ now has spacing $\omega$. It needs to be 
emphasized that this formal algebraic approach needs careful implementation, considering
the required square-integrability of the wave-functions.
\paragraph*{}For demonstrating the fact that, the scaling symmetry of the system directly
implies the underlying Heisenberg algebra, another set of similarity transformations 
leads to the Hamiltonian,

\bea
H_O&=&GFH_SF^{-1}G^{-1}\equiv-\frac{1}{2}\sum_i\frac{d^2}{dx_i^2}+\frac{\omega^2}{2}\sum_ix_i^2;\label{Sho1}\\
G&=&\exp\left(-\frac{\omega}{2}\sum_ix_i^2\right),~~~F=\exp\left(-\frac{1}{4\omega}\sum_i\frac{d^2}{dx_i^2}\right),\nonumber
\eea
representing $N$ decoupled oscillator modes of {\it Universal} frequency $\omega$.

\begin{figure}
\centering 
\includegraphics[width=3 in]{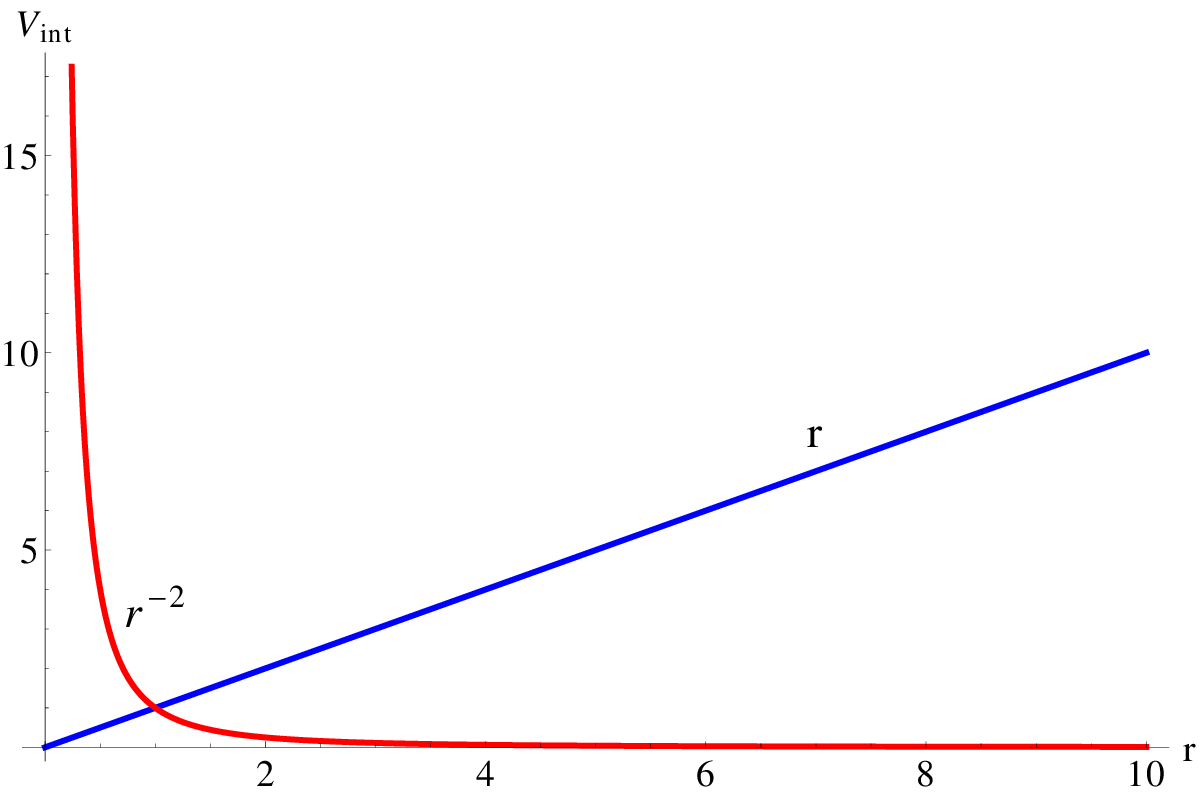}
\caption{\label{rev} Plots of solutions of the Laplaces equation in one dimension in comparison with $r^{-2}$ interaction.}
\end{figure}

\paragraph*{}As an explicit example with exact scaling symmetry, and to illustrate the exact role of
ground-state correlation, thereby defining the Hilbert space of the system, we consider the short-range
Calogero-Sutherland potential \cite{calogero},

\be
V_{CS}(\{x_i\})=\frac{g^2}{2}\sum^N_{i>j}\frac{1}{\vert x_i-x_j\vert^2}.\label{CSP}
\ee
It is worth noting that a systematic many-body treatment has generated this type of interaction in the
density functional approach \cite{DFCS}. An $x^{-2}$ potential is \textit{relatively} shorter than the Coulomb
interaction in 1-D, the latter being linear in $x$ (Fig. \ref{rev}). In generic $d$ dimensions, any
interaction with range shorter than $x^{-d}$ can be considered to be short ranged \cite{Manas}.
\paragraph*{}In the case of $V_{CS}$, the ground-state wave-function of the total Hamiltonian
$H_T$ takes the form \cite{Pitaevskii,calogero},

\bea
\Psi_0(\{x_i\})&\equiv&\prod_{i>j} \vert x_i - x_j\vert^{\alpha}\exp\left(-\frac{\omega^2}{2}\sum_kx_k^2\right),\label{JG}\\
\alpha&=&\frac{1}{2}\left(1 + \sqrt{1 + 4g^2}\right),\nonumber
\eea
of the Laughlin-type \cite{Laughlin}, 
where the Jastrow-type contribution $\psi_0=\prod_{i>j} \vert x_i - x_j\vert^{\alpha}$ \cite{N4},
represents the ground-state of the transformed Hamiltonian $\bar{H}$. The latter uniquely represents
the non-trivial pairing correlations. The Jastrow function,
with scaling `degree' $\alpha N(N-1)/2$, is known to be non-singular in the dimerization regime and
was realized earlier \cite{calogero,pkp1}, as a part of the exact solution for many-particle %confined
systems with $x^{-2}$ interactions.
\paragraph*{}For the sake of completeness, we note that for molecule formation, 
the 1-D unitary three-body problem in a trap can be modeled by the following contact conditions
on the wave-function,

\begin{equation} 
\psi(x_1,x_2,x_3) =\left(\frac{1}{x_{ij}}-\frac{1}{a}\right)A(X_{ij},r_{k})+\mathcal{O}(x_{ij}),\nonumber
\label{contact}
\end{equation}
in the limit $x_{ij}\equiv |x_i-x_j|\rightarrow 0$ taken for fixed positions of the other
particle $k$ and of the center of mass $X_{ij}$ of $i$ and $j$. For $A \neq 0$, the
wave-function is singular at $x_{ij}=0$, yet normalizable, suggesting molecule formation. However,
for the class of wave-functions in Eq. \ref{JG} representing short-range interaction, $A =0$,
exclusively marking dimerization. The wave-function is now well-behaved everywhere,
including $x_{ij} \rightarrow 0$, where it vanishes, much like the non-interacting fermions
or hard-core bosons. The density of states here is high, owing to the Heisenberg degeneracy. 
\paragraph*{}The Heisenberg symmetry of the C-S system can be extracted through transforming
the total Hamiltonian with respect to $\Psi_0$, yielding \cite{pkp1},

\be 
\tilde{H}= -\frac{1}{2}\sum_{i}\frac{d^2}{dx_i^2} - \alpha \sum_{i\ne j}\frac{x_i - x_j}{\vert x_i - x_j\vert^2}\frac{d}{dx_i}-\epsilon_0,\label{CS1}
\ee
with ground-state energy $\epsilon_0$ \cite{pkp1,pkp_mohanty} that can be expressed as,

\be
\epsilon_0=\left[\frac{1}{2}\alpha N(N-1)+\frac{1}{2}N\right]\omega.\label{GSE}
\ee
On comparison with Eq. \ref{F1}, in 1-D, the scaling parameter of Eq. \ref{scalingregime}
can now be identified as: $\gamma=\alpha N(N-1)/2$ for the C-S model \cite{pkp1}. The
identification $\sum_{i<j}n_in_j=N(N-1)/2$ \cite{Shankar} ensures inter-particle
correlation as the origin of this shift, under scaling symmetry. 
\paragraph*{}The Heisenberg symmetry in the C-S system was directly
illustrated, through further similarity transformations, yielding \cite{pkp1},

\be\label{HD}
H_D=\frac{\omega}{2}\sum_i\left\{a_i^-,a_i^+\right\}+\epsilon_0-\frac{\omega}{2}N,
\ee
with suitable definitions of the operators $a_i^\pm$. %This can also be achieved by a unitary transformation 
%of operators \cite{Gonera}. 
The system can be expressed as 
$N$ non-interacting oscillators of frequency $\omega$, with exact shift 
$\gamma\hbar\omega$ in the ground-state energy \cite{pkp1}, which was \textit{not} observed in
earlier works \cite{Werner,Pitaevskii}. With the ground-state energy removed, through similarity
transformations, the first term on the RHS represents {\it excited} states, spaced by $\omega$.
Further, overlooking spin degeneracy, the Fermi energy of $N$ non-interacting fermions in a
harmonic trap, having spectrum given by Eq. \ref{HD}
is $\epsilon_F=N\omega$. Thus, for C-S model, $\epsilon_0=\xi\epsilon_F$,
with Universal scaling $\xi=(1/2)\left[1+\alpha(N-1)\right]$, similar to Ref. \cite{three}
in 3-D. This is also expected from the Universal proportionality of C-S partition funtion
with the N-fermion one \cite{Shankar}.
\paragraph*{}In the absence of a trap, equivalence between fermions, with interaction of the C-S type,
and decoupled ones can be achieved by making use of {\it two} different $SU(1,1)$ algebras, 

\bea
[T_+,T_-]&=&-2\omega T_0\quad,\quad [ T_0 , T_{\pm} ] = \pm T_{\pm};\label{su11}\\
T_0&=& -\frac{1}{2}\sum_iD_i-\frac{\epsilon_0}{2},~~~T_-=\frac{\omega}{2}\sum_i x_i^2,\nonumber\\
T_+&=&\left\{\begin{array}{rl}
&\frac{1}{2}\sum_{i}\frac{d^2}{dx_i^2}+\alpha\sum_{i>j}\frac{1}{x_i - x_j}\frac{d}{dx_i}\\
&\frac{1}{2}\sum_i\frac{d^2}{dx_i^2}
\end{array} \right.,\nonumber
\eea
corresponding to interacting and non-interacting systems, respectively. This owes to the special property of
the Euler operator, allowing combination of operators, scaling with same degree,
to form a new algebra. The above generators obtained, after performing certain similarity
transformations, are tailor made for the class of wave-functions of Eq. \ref{JG} in the scaling
regime and are slightly different from Ref. \cite{Pitaevskii} in form. In particular, our algebra is
valid near the infinite scattering limit, where the scaling law of Eq. \ref{scalingregime} comes in to effect. 
\paragraph*{}The C-S Hamiltonian {\it without} any confinement,

\be
H_{CS}=-\frac{1}{2}\sum_i\frac{d^2}{dx_i^2}+\frac{g^2}{2}\sum_{i>j}\frac{1}{\vert x_i-x_j\vert^2},\nonumber
\ee
can be mapped to a free system \cite{Gonera}, through the following
similarity transformation,

\bea
H_{\rm free}&=&e^{\eta\sum_ix_i^2}e^{\beta{\cal B}}\Psi_0^{-1}H_{CS}\Psi_0e^{-\beta{\cal B}}e^{-\eta\sum_ix_i^2}\nonumber\\
&\equiv&\mp\sqrt{1+\omega^2}\left(\sum_iD_i+\epsilon_0\right);\label{Free}\\
{\cal B}&:=&\frac{1}{2}\sum_i\frac{d^2}{dx_i^2}+\alpha\sum_{i\ne j}\frac{1}{x_i-x_j}\frac{d}{dx_i},\nonumber\\
\beta&=&\frac{1}{\omega^2}\left(1\pm\sqrt{1+\omega^2}\right),~~\eta=\pm\frac{\omega^2}{4\sqrt{1+\omega^2}}.\nonumber
\eea 
Here $\tau=1/\omega$ defines a suitable time-scale for the system, inherent to C-S model with underlying 
$SO(2,1)$ symmetry of Eq. \ref{su11}, with or without the Harmonic trap. This shows that the shift
$\gamma/\tau$, in $\epsilon_0$, is purely due to $N$-particle correlation. %It further illustrates
\paragraph*{}It is to be noted that the wave-functions of this `free' Hamiltonian are highly 
correlated ones. This system
obeys fractional statistics~\cite{Polychronakos}, and more specifically, fractional exclusion
statistics \cite{Bhaduri,Baskaran,FracStatPKP}. 
Universality of the degenerate gas on the other hand, suggests that the scaling exponent
$\gamma$ is independent of the specific statistics of the system. Nevertheless, at unitarity,
it might be useful to model the interaction as coming from fractional exclusion statistics and
intriguingly, the estimates for energy per particle are in good agreement with Monte-Carlo
simulations \cite{Bhaduri}. 
%%%%%%%%%%%%%%%%%%%%%%%%%%%%%%%%%%%%%%%%%%%%%%%%%%%%%%%%%%%%%%%%%%%
\paragraph*{}This Universal shift of the ground state energy relates well with
the recent correspondence between non-relativistic conformal systems and their gravity duals
\cite{Son:2008ye,Balasubramanian:2008dm}. Each value of the scaling exponent is in general
related to the dimension of primary operators ~\cite{Son:2008ye,Nishida:2007pj}. For two particles of 
opposite spins, $\gamma$ can take values $0$ or $-1$~\cite{Stan}, which means $\alpha$ takes
exactly the same values. For the case of three particles $\gamma \approx 0.22728$ and
$\alpha=0.07576$. Since, both the $SU(1,1)$ algebras in Eq. \ref{su11} can be embedded
in the Schr\"odinger algebra, the exact map we constructed, confirms the duality between free and
unitary fermions envisaged in~\cite{Son:2008ye}, which was originally argued based on
insights from the AdS/CFT correspondence~\cite{Klebanov:1999tb} regarding the  existence of a
pair of non relativistic conformal field theories with operators of different dimensions. These features
continue to hold for the C-S type of models, as emphasized in~\cite{Wen:2008hi}. Further, this
duality is general to any short-range potential, in appropriate dimensions, satisfying
Eq. \ref{scalingregime}. 
\paragraph*{}In conclusion, it has been shown that, an one-dimensional N-body interacting
system at unitarity, with or without trap, exactly maps to a non-interacting one, respecting Heisenberg algebra.
The generator of the transformation being the scaling operator, it is necessary that the
system is scale-covariant. This corresponds to homogeneous polynomial wave-functions, exclusively
representing the dimerization regime. The underlying algebra establishes the existence of an
$\omega$ breathing mode in 1-D, not observed in earlier literature. Also, the shift in the ground
state energy due to scaling properties, represents the role of correlation in 1-D. Further, the
possibility of exclusion statistics of the fermions near unitarity is pointed out, which is due to the $x^{-2}$ type
short-range interaction, supported by Monte Carlo simulations.
 
{\it{Acknowledgements:}} CB would like to thank IISER Kolkata and Chiranjib Mitra for wonderful hospitality
and Kuntala Bhattacharjee for useful conversations.

\end{document}